\documentstyle{article}
\begin{document}

\title{%
\vskip-6pt \hfill {\rm\normalsize UCLA/01/TEP/6} \\  
\vskip-6pt \hfill {\rm\normalsize  August 2001} \\
\vskip-9pt~\\
Expected Signals in Relic Neutrino Detectors}

\author{Gintaras Duda and Graciela Gelmini \\
Department of Physics and Astronomy, UCLA \\ Los Angeles, CA 90095-1547 \\
\and
Shmuel Nussinov \\ Tel Aviv University, Ramat Aviv, Tel Aviv, Israel}

\maketitle

\begin{abstract}

Here we estimate the magnitude of the signals expected for realistic cosmic
neutrino backgrounds in detectors attempting to measure the mechanical
forces exerted on macroscopic targets by the elastic scattering of relic
neutrinos.  We study effects proportional to the weak coupling constant
$G_F$ and to $G_F^2$ for Dirac and Majorana neutrinos, either relativistic
or non-relativistic, both gravitationally bound or not.
\end{abstract}  

\section{Introduction}

Alongside the  well measured background
of cosmic photons, the Big Bang model  predicts the existence of an elusive
background of cosmic neutrinos.  This neutrino background
may  be revealed in the near future by measurements of the microwave 
background radiation anisotropy and the large-scale structure of the Universe.

Here we reconsider the 
long-standing question of the detectability of the neutrino background in a
laboratory experiment, a recurrent subject in the literature for the past 
thirty years \cite{Stodolsky}-\cite{Hagmann}.  
We estimate how far macroscopic accelerations due to realistic neutrino 
backgrounds are from the smallest measurable acceleration at present, which is
about $10^{-12}$ cm/sec$^2$\cite{minacc}.  The new elements we consider in our 
estimates are the possibility of a very large lepton asymmetry in the cosmic
neutrino background, and the study of Majorana as well as Dirac neutrinos.

With no lepton number asymmetry the number density $n_{\nu_i}$ of the 
neutrinos
of each species in the background is equal to the number of anti-neutrinos of 
each species.
The Big Bang model predicts this number density to be $n_\nu = (3/22)~n_\gamma = 
0.136~n_\gamma$,
where $n_\gamma$ is the density of the microwave background radiation photons.
Since $n_\gamma = 412$ cm$^{-3}$, therefore $n_{\nu_i} = 56$ cm$^{-3}$.  It is,
however, possible for neutrinos to have very large lepton asymmetries.  While
charge neutrality requires the asymmetry in charged leptons to be the same
as that in protons, for which ($n_B-n_{\bar B})/n_\gamma \simeq O(10^{-10})$, 
no
such requirement limits the asymmetry in neutrinos.  With a large asymmetry
between neutrinos and anti-neutrinos only the excess of 
either $\nu_i$ or  $\bar\nu_i$ remains after annihilation ceases  with a
density
\begin{equation}
n_{\nu_i} = n_\gamma~0.0252
\left[\xi_i \pi^2 +
\xi_i^3\right]
\label{eq:1} 
\end{equation}
where $\xi_i= |\mu_{\nu_i}|/T_{\nu_i}$ and $\mu_{\nu_i}$ is the 
chemical potential of the given neutrino species. The ratio
$\xi_i$ is constant after neutrinos decouple.

Thus, for example, with $\xi_i= 5$ \cite{KS,AS,LP},
one obtains a neutrino density of
$n_\nu \simeq 4n_\gamma = 1700$ cm$^{-3}$.  Eq. (1) is valid for
$\xi_i < 12$ \cite{KS} (for which the decoupling temperature
$T_{dec}$ is smaller than $2m_\mu$, see below). Bounds coming from Big Bang 
Nucleosynthesis (BBN), the anisotropy of the microwave background radiation and
data on the large scale structure of the Universe limit 
$\xi_i$ to
be much lower than 12.  Kang and Steigman in 1992 \cite{KS} obtained
\begin{equation}
-0.06 \leq \xi_e\leq 1.1~, ~~~~
\left|\xi_{\mu,\tau}\right| \leq 6.9~.
\label{eq:2} 
\end{equation}
More recent studies yield bounds that depend on the presence of a cosmological
constant (with energy density $\Omega_\Lambda$ in units of the critical 
density).
If the cosmological constant is zero or small, $\Omega_\Lambda < 0.1$, the
value $\xi_i \simeq 5$, namely $n_{\nu_i}/n_\gamma \simeq 4$ is 
favored
\cite{AS},\cite{LP}. 
 For $\Omega_\Lambda \simeq 0.5$, one obtains $n_{\nu_i}/n_\gamma < 4$, and for 
 $\Omega_\Lambda > 0.7$, the lepton
asymmetry must be very small, so the neutrino density goes back to the standard 
value,
$n_{\nu_i}/n_\gamma \simeq 0.14$ \cite{AS, KR}. Notice that in models
with large neutrino asymmetries, the  standard BBN mechanism is
altered, allowing for the existence of many more neutrinos than in 
the standard case.  BBN and recent cosmic microwave background anisotropy
data provide constraints on neutrino asymmetries which depend strongly on
cosmological parameters \cite{S2}.  In what follows
we will take an upper bound $n_{\nu_i}/n_\gamma \leq 4$, i.e. $n_{\nu_i}\leq
1700$ cm$^{-3}$, when considering the case of large lepton asymmetries.

The decoupling temperature of neutrinos $T_{\rm dec}$, 
which with no lepton asymmetry is a few MeV, increases with the
lepton asymmetry. However, only for $\mu_{\nu_i}/T_{\nu_i} \geq 12$ would 
$T_{\rm dec}$ become larger 
than 2$m_{\mu}$, i.e. neutrinos would decouple before
$\mu^+\mu^-$ annihilation. In this case, not only would $e^+e^-$ annihilations
increase the number of photons relative to that
of neutrinos due to entropy conservation, but  
$\mu^+\mu^-$ annihilations would as well, leading to a lower neutrino 
temperature relative to the photon temperature. 
For $\mu_{\nu_i}/T_{\nu_i} < 12$ the temperature of $\nu_i$ is now the usual 
one,
$T_{\nu_i} = (4/11)^{1/3}~T_\gamma = 0.71~T_\gamma$. 
With $T_\gamma = 2.728^\circ$K, this means
$T_{\nu_i} = 1.95^\circ$K = 1.68~$10^{-4}$ eV.
 Therefore, if $m_{\nu_i}$ is smaller than
$1.7~10^{-4}$ 
eV background neutrinos  are relativistic at present while for larger masses
they are non-relativistic. Moreover neutrinos may be Dirac or Majorana
particles, a distinction that is important only for non-relativistic
neutrinos.

Relativistic neutrinos are only in left-handed chirality states (and
anti- neutrinos only in right-handed chirality states). These are  
the only states produced
by weak interactions.   For relativistic neutrinos chirality and helicity 
coincide (up to 
mixing
terms of order $m_{\nu_i}/ E_\nu \simeq m_{\nu_i}/T_\nu$). 
Bounds from structure formation in the Universe imply that stable neutrino 
masses are at most of the order of a few eV \cite{primack}.  This means that
neutrinos were relativistic at decoupling $(T_{\rm dec} \geq O({\rm MeV})$),
even if they may be non-relativistic at present.  

We will call left(right)
chirality eigenstates $\nu_L(\nu_R)$ and left(right) helicity eigenstates
$\nu_{\ell}(\nu_r)$. At decoupling, neutrinos $\nu_L$ were
only in left-handed helicity states and antineutrinos $\nu_R^c$ (or
$\nu_R$ in the case of Majorana neutrinos) in right-handed ones.
Helicity is an eigenstate of
propagation and therefore it does not change while neutrinos propagate freely,
even if they become non-relativistic.  For Majorana neutrinos chirality acts
as lepton number, so we are calling ``neutrinos" those particles produced
at $T>T_{dec}$ as $\nu_L$ and ``anti-neutrinos" those produced as $\nu_R$.
Thus, neglecting intervening interactions, non-relativistic background neutrinos 
are
in left-handed helicity eigenstates (which consist of equal admixtures of
left- and right-handed chiralities) and anti-neutrinos are in right-handed
helicity eigenstates (which also consist of equal admixtures of
left- and right-handed chiralities). 
If the non-relativistic neutrinos are Dirac particles, only the left-handed
chirality states (right for anti-neutrinos) interact, since the  other 
chirality state is sterile, while if the
neutrinos are Majorana, both chirality states interact (the right-handed
``neutrino" state is the right-handed anti-neutrino).

Slow enough non-relativistic neutrinos eventually fall into gravitational 
potential wells, become bound and,
 after a characteristic orbital time, their helicities 
become well mixed up, since momenta are reversed and  spins are not.  
Thus, gravitationally bound background neutrinos  have well mixed helicities. 
Most background neutrinos however are not gravitationally bound at present because 
they are too light.

The present upper bound from structure formation in the Universe for neutrinos 
without a lepton  asymmetry, for which $n_\nu \simeq 100$ cm$^{-3}$, is of a few eV
\cite{primack}.
 Neutrinos carrying a large lepton asymmetry must necessarily be lighter. 
 The upper bound on the mass of degenerate neutrinos with a large asymmetry should be 
lower, since their number density is larger, at most of the order of 0.1 eV.
 The Tremaine-Gunn kinematical constraint 
\cite{tremaine-gunn} gives the scale at which neutrinos gravitationally cluster.  Light 
neutrinos with masses $m_\nu < eV$ would be 
gravitationally bound only to the largest structures, large
clusters of galaxies. We can see this  using simple velocity arguments.
 Only cosmic neutrinos with velocities smaller
 than the escape velocity of a given structure can be bound to it.
The escape velocity from a large galaxy 
like ours is about 600 km/s and from a large cluster of galaxies is about 2,000 km/s.
Considering that the average velocity modulus  of  non-relativistic neutrinos
 of mass $m$ and  temperature  $T_\nu$ is  (using Maxwell-Bolztman distribution)
 $\langle|\vec\beta_\nu|\rangle$=$\sqrt{8kT_\nu/ \pi m}$=$\sqrt{4.3 ~10^{-4} eV/m}$
 (namely $\langle|\vec v_\nu|\rangle$= 6,200 km/s for $m=1$eV, 
 and $\langle|\vec v_\nu|\rangle$= 19,600 km/s for $m=0.1$eV) 
it is obvious that only about a third of 1 eV mass neutrinos and  a very small
fraction of lighter neutrinos  could be gravitationaly bound to large clusters at present.
Fermi degenerate neutrinos may have even larger average velocities depending on their
chemical potential $\xi=\mu/T_\nu$ (constant after neutrinos decouple), 
but the conclusions remain the same. For  $\xi>>1$
 $\langle|\vec\beta_\nu|\rangle$=$\sqrt{6 \xi T_\nu/ 5 m}
  \simeq \sqrt{ \xi 1.68~10^{-4} eV/m}$ (namely 
  $\langle|\vec v_\nu|\rangle$= $\sqrt{\xi}$ 12,300 km/s for $m=0.1$eV) and both expressions 
  coincide for $\xi= 2.5$.
In all cases the amount of neutrinos in the tail of the velocity distribution with
velocities 
smaller that 600km/s which would be gravitationally bound to galaxies is much smaller.

In the following we will give our results for both clustered (C-NR) and non clustered
(NC-NR) non-relativistic background neutrinos as well as for relativistic (R) 
background neutrinos.

With zero lepton asymmetry the background consists of equal numbers of neutrinos and
anti-neutrinos. Thus we have equal numbers of left and right either chirality or
helicity states in all cases, relativistic or non-relativistic, Dirac or
 Majorana neutrinos.  In the case of a large lepton
asymmetry which favors, say, neutrinos so that there are only $\nu_L$ at the 
moment neutrinos decouple, 
(left and right  are to be exchanged in the following argument for an 
asymmetry favoring anti-neutrinos) the background consists of left chirality 
particles in the case of R or NC-NR neutrinos, and it consists of
totally mixed helicity states for C-NR neutrinos

 These properties of the different
possible neutrino backgrounds are relevant to the  effect linear in the weak
coupling constant $G_F$, first studied by Stodolsky \cite{Stodolsky}.

\section{The $G_F$ Effect}

This effect was proposed by Stodolsky in 1974 (before  weak
neutral currents were proven to exist, so he did not include them). 
It is the energy split of the two spin states of 
non-relativistic electrons in the cosmic neutrino bath.   This energy split is
proportional to the difference between the densities of neutrinos and 
antineutrinos in the 
neutrino
background 
for Dirac neutrinos, and 
proportional to the net helicity of the
background for  Majorana neutrinos, as we will now see.

The Hamiltonian density of the $\nu$-$e$ interaction is 
\begin{equation}
{\cal{H}}(x) = 
\frac{G_F}{\sqrt{2}}~ \bar e\gamma^\mu(g_V - g_A \gamma_5)e~
\bar\nu \gamma_\mu (1 - \gamma_5)\nu~.
\label{eq:3} 
\end{equation}
For $\nu =\nu_e$, $g_A= 1/2$ and $g_V= 1/2 + 2 \sin^2{\theta_W}$ while
for $\nu= \nu_\mu$ or $\nu =\nu_\tau$, with which electrons only have neutral 
weak
interactions,  
$g_A= -1/2$ and $g_V= -1/2 + 2 \sin^2{\theta_W}$.

\subsection{Dirac Neutrinos}

We compute first the energy shift of a single electron of momentum $p$ and
spin $s$ in the neutrino background: $\Delta E_e = \langle p,s |H|p,s\rangle$.
Let us call $p$ and $k$ the momenta of the $e$ and $\nu$ respectively, 
and $s(s^\prime)$ the incoming(outgoing) spins.  Working in momentum space
and using the second quantization form for the neutrino fields, we obtain for
the case of Dirac neutrinos

\begin{eqnarray}
\left<H^D\right> & = & \frac{G_F}{\sqrt{2}}\frac{m_e}{E_eV}
\left(\sum_{s,s^\prime} \bar u_e(p,s^\prime)\gamma^\mu
(g_V-g_A \gamma_5) u_e(p,s)\right) \nonumber \\
& \times & \left\{\int d^3k\left(\frac{m_\nu}{E_\nu}\right)
\sum_{s,s^\prime} b^{\dagger}(k,s^\prime) b(k,s)
\left[ \bar u_\nu(k,s^\prime)
\gamma_\mu(1-\gamma_5)u_\nu(k,s) \right]\right. \nonumber \\
& - &\left.\int d^3k\left(\frac{m_{\nu^c}}{E_{\nu^c}}\right)
\sum_{s,s^\prime} d^{\dagger}(k,s^\prime) d(k,s)
\left[\bar v_\nu(k,s)\gamma_\mu(1-\gamma_5)
v_\nu(k,s^\prime)\right]\right\}
\label{eq:4} 
\end{eqnarray}
where $u$ and $v$ are the usual spinors for particles and antiparticles
and $b$, $d$ ($b^{\dagger}$, $d^{\dagger}$) are the respective annihilation 
(creation) operators, for neutrinos and antineutrinos respectively, of momentum 
$k$.  
In the rest frame of the electron ($p=0$) the diagonal matrix element of
the weak current is $g_A \sigma^i_e$.  In this frame the neutrino current
is $J^i_{\nu}=-\beta^i_{earth}\left(n_\nu-n_{\nu^c}\right)$, since 
the non-zero average velocity  $-\vec\beta_{earth}$ of the neutrinos 
is due to the motion of the 
earth relative to the neutrino bath (with velocity 
  $\vec\beta_{earth}$).  Dotting the two together yields
 $\Delta E = - (G_F/\sqrt{2}) g_A \left(\vec\sigma_e \cdot \vec\beta_{earth}\right)
\left(n_\nu-n_{\nu^c}\right)$.  

We would like to present however a more careful,
systematic, and first principle derivation.  Using the decomposition
$u_e^T(p,s)$ = $\sqrt{(E_e+m_e)/2m_e}$ $\left(\chi^T_s,~~[(\vec\sigma_e\cdot\vec 
p) / (E_e + m_e)]\chi_s^T \right)$ for the electron spinors and the Gordon 
decomposition and similar relations containing $\gamma_5$ for the neutrino 
vertices, one
obtains
\begin{eqnarray}
\left<H^D\right> &=& \frac{G_F}{\sqrt{2}} \left(\frac{E_e+m_e}{E_e}\right)
\left\{g_V\left[1 + \left(\frac{2\vec\sigma_e\cdot\vec 
p}{E_e+m_e}\right)^2\right] 
- 2g_A\frac{2\vec\sigma_e\cdot\vec p}{E_e+m_e}~{\bf ,}\right. \nonumber \\
& &{} \left.2\vec\sigma_e\left(2g_V\frac{2\vec\sigma_e\cdot\vec 
p}{E_e+m_e}-g_A\left[1 
+ \left(\frac{2\vec\sigma_e\cdot\vec p}{E_e+m_e}\right)^2\right]\right)\right\} 
\nonumber \\
& &{} \times \frac{1}{V}\left\{\int d^3k
\sum_{s,s^\prime} b^{\dagger}(k,s^\prime) b(k,s)
\frac{1}{E_\nu}
\left[\bar u_\nu(k,s^\prime)(k_\mu-2W_\mu) u_\nu(k,s)\right]\right.
\nonumber \\
& &{} + \left.\int d^3k
\sum_{s,s^\prime} d^{\dagger}(k,s^\prime) d(k,s)
\frac{1}{E_{\nu^c}}
\left[\bar v_\nu(k,s)(k_\mu-2W_\mu)
v_\nu(k,s^\prime)\right]\right\}
\label{eq:5} 
\end{eqnarray}
Here the comma in the first key bracket separates the zero and spatial
components of the electron amplitude and $W_\mu =i \sigma_{\mu \nu} k^\nu 
\gamma_5 =-\frac{1}{2}\epsilon_{\mu\nu\sigma\rho}\sigma^{\sigma\rho}k^\nu$  
is the Pauli-Lubansky pseudo-vector (one  uses 
here $\gamma_5\sigma_{\mu\nu} =
\frac{i}{2}\epsilon_{\mu\nu\sigma\rho}\sigma^{\sigma\rho}$). In terms of the 
four
dimensional ``spin operator'' 
\begin{equation}
\vec\Sigma=\left(\begin{array}{cc} \vec\sigma & 0 \\
0 & \vec\sigma 
\end{array}\right)
\label{eq:199}
\end{equation}
the Pauli-Lubansky  pseudo-vector is given by
\begin{equation}
W_\mu = \left(\frac{\vec\Sigma}{2}\cdot\vec k~{\bf ,} ~~
-\frac{E_\nu}{2}\vec\Sigma + \frac{i}{2} \gamma_5 (\vec k\times 
\vec\Sigma)\right)~,
\label{eq:6} 
\end{equation}
so that for particles at rest $(\vec k=0)$ it is proportional to the spin 
operator
$W_\mu = (0,~-m_\nu \vec \Sigma/2)$ and for relativistic particles it is 
proportional to the helicity $h=\vec \Sigma\cdot\hat k$, i.e. $W_\mu|k\rangle 
= (h/2)k_\mu|k\rangle$.  Although $\vec W$ is the spin operator in the rest 
frame and 
$W^{\mu}W_{\mu}=-m^{2}s(s+1)$
is a Casimir operator of the Poincare algebra, in general $\vec W$ is not the 
spin operator: 
it is easy to see that
the $W_i$ do not have the SU(2) commutation relations required for angular 
momentum 
components, 
i.e. $[W_i,W_j]\not=i\epsilon_{ijk}W_k$.
However, one can write $\vec W$ in terms of the true spin operator 
\cite{GurnseyDeWitt}
\begin{equation}
\vec S = \frac{1}{m_\nu}\left(\vec W - \frac{W_{0}\vec k}{E_\nu
+ m_\nu}\right)
\label{eq:198}
\end{equation}
Considering non-relativistic electrons with velocity $\vec\beta_e= \vec p/ 
E_e$ (thus $(E_e+m_e )/ E_e= 2$),
using $\vec\beta_\nu = \vec k/ E_\nu$, and writing $\vec W$ in terms of $\vec 
S$, we obtain 
to first order in 
$\vec\beta_e$
\begin{eqnarray}
\left<H^D\right> &=& \frac{G_F}{\sqrt{2}}\left\{(g_V-g_A ~2\vec 
s_e\cdot\vec\beta_e),~~
2\vec s_e(-g_A+g_V ~2\vec s_e\cdot\vec\beta_e)\right\} \nonumber \\
& &{}\frac{1}{V}\left\{\int d^3k
\sum_{s,s^\prime} b^{\dagger}(k,s^\prime) b(k,s)
\left[\bar u_\nu(k,s^\prime)
\left(1-\vec \Sigma\cdot\vec\beta_\nu~,\right. \right. \right. \nonumber \\
& &{}\left.\left.\left. -\vec\beta_\nu+\frac {2m_\nu}{E_\nu}\vec S + \frac 
{E_{\nu}\vec\beta_\nu}{E_{\nu}+m_{\nu}}\vec
\Sigma\cdot\vec\beta_\nu\right) 
 u_\nu(k,s)\right]
\right. \nonumber \\
& &{} \left. +\int d^3k 
\sum_{s,s^\prime} d^{\dagger}(k,s^\prime) d(k,s)
\left[\bar v_\nu(k,s)
\left(1-\vec \Sigma\cdot\vec\beta_\nu~,\right.\right.\right. \nonumber \\
& &{}\left.\left.\left. -\vec\beta_\nu+\frac {2m_\nu}{E_\nu}\vec S + \frac 
{E_{\nu}\vec\beta_\nu}{E_{\nu}+m_{\nu}}\vec
\Sigma\cdot\vec\beta_\nu\right) v_\nu(k,s^\prime)\right] \right\}~.
\label{eq:7} 
\end{eqnarray}
In the neutrino ``rest-frame'' the term of the form $\vec s_e \cdot\vec S$ in 
Eq.~(\ref{eq:7}) 
vanishes for the following simple reason.
We have calculated our interaction between a ``sea'' of neutrinos and a single 
electron.
Thus $\vec s_e\cdot\vec S$ gives the projection of the individual neutrino spin 
along the 
single electron's
spin axis.  But we expect the neutrinos to be distributed isotropically, and 
thus there is 
equal probability for the
neutrino spin projected along this axis to be positive or negative.  Hence when 
the sum over 
spins and the integration
over the isotropic neutrino distribution are performed, these terms becomes 
zero.  However, 
in the electron rest-frame
the neutrino distribution is no longer isotropic and this term does not vanish.  
Using the 
basis of neutrinos of 
definite helicity, we can take the spin to lie completelty parallel or 
anti-parallel to the 
direction of motion, i.e. 
$\vec S = (h/2)~\hat\beta$ where h is the helicity.  Thus the term in 
Eq.~(\ref{eq:7}) 
containing the neutrino spin will 
be proportional to $\langle (m_\nu / E_\nu)~\hat\beta\rangle$.

Instead of utilizing the spin basis, we now work in the helicity basis so that 
all spin dependent terms involve the helicity operator $\vec\Sigma\cdot\hat\beta$. We 
can use now the completeness relations for the spinors $u$ and $v$,
\begin{equation}
\bar u_\nu(k,h^\prime) u_\nu(k,h) =\delta_{h h^\prime}~~~~
\bar v_\nu(k,h^\prime) v_\nu(k,h) =-\delta_{h h^\prime}.
\label{eq:8} 
\end{equation}
Furthermore, in the helicity basis we have
\begin{equation}
\bar u_\nu(k,h^\prime) ~\vec \Sigma_\nu \cdot\vec\beta_\nu ~u_\nu(k,h)= h 
|\vec\beta_{\nu}|~
 \delta_{h h^\prime},
\label{eq:11} 
\end{equation}
\begin{equation}
\bar v_\nu(k,h^\prime) ~\vec \Sigma_\nu \cdot\vec\beta_\nu~ v_\nu(k,h)= -h 
|\vec\beta_{\bar\nu}|~
 \delta_{h h^\prime},
\label{eq:12} 
\end{equation}

where $h= 2 \vec s \cdot \hat\beta$ is the helicity. Thus we obtain
\begin{eqnarray}
&&\left<H^D\right> = \frac{G_F}{\sqrt{2}}\frac{1}{V}\left\{(g_V-g_A ~2\vec 
s_e\cdot\vec\beta_e)~ {\bf ,}~~
2\vec s_e(-g_A+g_V ~2\vec s_e\cdot\vec\beta_e)\right\} \times \nonumber \\
&&{} \int d^3k \left\{
\sum_{h} b^{\dagger}(k,h) b(k,h)
\left(1-h,~-\vec\beta_\nu+\frac{E_{\nu}h}{E_\nu+m_\nu}|\vec\beta_\nu|~\vec\beta_
\nu
+ \frac{m_{\nu}}{E_\nu}h~\hat\beta_\nu\right) 
\right. \nonumber \\
&&{} \left.-
\sum_{h} d^{\dagger}(k,h) d(k,h)
\left(1-h,~-\vec\beta_\nu+\frac{E_{\nu}h}{E_\nu+m_\nu}|\vec\beta_\nu|~\vec\beta_
\nu
+ \frac{m_{\nu}}{E_\nu}h~\hat\beta_\nu\right)\right\}.
\label{eq:13} 
\end{eqnarray}

The number density of neutrinos  and antineutrinos  are respectively
\begin{equation}
n_{\nu} = \frac{1}{V}\int d^3k \sum_{h} b^{\dagger}(k,h) b(k,h),
\end{equation}
\begin{equation}
n_{\nu^c} = \frac{1}{V}\int d^3k \sum_{h} d^{\dagger}(k,h) d(k,h),
\end{equation}
and obviously
\begin{equation}
\frac{1}{V}\int d^3k \sum_{h} b^{\dagger}(k,h) b(k,h) \vec\beta_\nu =
\langle\vec\beta_\nu\rangle n_\nu~,
\label{eq:14} 
\end{equation}
and similarly for antineutrinos, where $\langle\ \rangle$ denotes average 
values.

Recall that we denote left- and right-handed helicity states with
lower case indices, $\nu_{\ell}$ and $\nu_r$.  In general the neutrino and the 
anti-neutrino are admixtures of states of left and right helicity.
Thus we write
\begin{equation}
n_{\nu} = n_{\nu_\ell} + n_{\nu_r}~, ~~
n_{\nu^c} = n_{\nu^c_\ell} + n_{\nu^c_r}~.
\label{eq:15} 
\end{equation}
With this notation the terms in $\left<H^D\right>$ linear in $\vec s_e$ are 
$H^D_{s_e}$,
\begin{eqnarray}
H^D_{s_e}&=&
(G_F / \sqrt{2})
\left\{-g_A~2\vec s_e\cdot\vec\beta_e
\left[(n_{\nu}-n_{\nu^c})+ \langle|\vec\beta_\nu|\rangle
(n_{\nu_\ell} + n_{\nu^c_\ell}-n_{\nu_r} - n_{\nu^c_r})\right]\right.\nonumber 
\\
&+&  g_A~2\vec s_e\cdot\left[\langle\vec\beta_\nu\rangle 
(n_{\nu}-n_{\nu^c})+\right.\nonumber \\
& &\left.\left.\left(\left< {\frac{E_\nu}{E_\nu+m_\nu} \vec\beta_\nu 
|\vec\beta_\nu|} \right>
+\left<\frac{m_\nu}{E_\nu}~\hat\beta_\nu \right>\right)
(n_{\nu_\ell} + n_{\nu^c_\ell}-n_{\nu_r} - n_{\nu^c_r})\right]
\right\}.~
\label{eq:16} 
\end{eqnarray}
In the rest frame of the $\nu$-bath due to isotropy 
the average of all vectors proportional to neutrino velocities
 $\langle\vec\beta_\nu\rangle=0$, 
$\langle\hat\beta_\nu\rangle=0$, 
$\langle(E_\nu/{E_\nu+m_\nu}) 
\vec\beta_\nu |\vec\beta_\nu|\rangle=0$ and $\langle (m_\nu 
/E_\nu)~\hat\beta_\nu   
\rangle=0$, thus only the 1st. term in Eq. (18) remains.
In the rest frame of the electron, the frame 
in which experiments are performed, $\vec\beta_e=0$ and only the 2nd. 
term in Eq. (18) remains. In this frame
 the non-zero value of 
$\langle\vec\beta_\nu\rangle$ is due to the motion of the earth relative
to the neutrino bath, $\langle\vec\beta_\nu\rangle=-\vec\beta_{earth}$,
 which is approximately $10^{-3}$.  
One can compute the averages in Eq. (\ref{eq:16}) explicitely in the 
relativistic (R) and 
non-relativistic (NR) limits, using a Gaussian distribution for the neutrinos in 
the latter 
case. Up to first order in $\vec\beta_{earth}$ we have
\begin{equation}
\left<\frac{E_\nu}{E_\nu+m_\nu}\vec\beta_\nu|\vec\beta_\nu|\right>_{R}=-
\langle|\vec\beta_\nu|\rangle\vec\beta_{earth}=-\vec\beta_{earth},
\label{eq:197}
\end{equation}
\begin{equation}
\left<\frac{m_\nu}{E_\nu}\hat\beta_\nu\right>_{R}=-\frac{\vec\beta_{earth}}
{\gamma},
\label{eq:657}
\end{equation}
\begin{equation}
\left<\frac{E_\nu}{E_\nu+m_\nu}\vec\beta_\nu|\vec\beta_\nu|\right>_{NR}=-\frac{2}{3}
\langle|\vec\beta_\nu|\rangle\vec\beta_{earth} + O(\vec\beta^3_{earth}),
\label{eq:196}
\end{equation}
\begin{equation}
\left<\frac{m_\nu}{E_\nu}\hat\beta_\nu\right>_{NR}=
-\frac{16}{3\pi}~\frac{\vec\beta_{earth}}{
\langle|\vec\beta_\nu|\rangle} 
+ O(\vec\beta^3_{earth}),
\label{eq:658}
\end{equation}
where $\langle|\vec\beta_\nu|\rangle$ is always the average of the velocity 
modulus in the 
neutrino rest frame. We see in Eqs. (19) and (21) that 
$\langle(E_\nu/{E_\nu+m_\nu}) \vec\beta_\nu |\vec\beta_\nu|\rangle$ is either of the 
same order of magnitude or smaller than $\vec\beta_{earth}$. 
The remaining average, $\langle (m_\nu /E_\nu)~\hat\beta_\nu   
\rangle$ is negligible for relativistic neutrinos with $\gamma >> 1$, but is 
larger than $\vec\beta_{earth}$ for non-relativistic neutrinos, for which
 $\langle|\vec\beta_\nu|\rangle< 1$ (see Eq. (22)). 
 The factor $(n_{\nu_\ell} + n_{\nu^c_\ell}-n_{\nu_r} - n_{\nu^c_r})$,
  which multiplies both these averages in Eq. (18), becomes zero for C-NR neutrinos
(since the helicities are well mixed) but could be large for NC-NR  neutrinos 
in the case of a large lepton asymmetry. In this case  
 $-\langle (m_\nu /E_\nu)~\hat\beta_\nu\rangle$ = 
 $1.7\sqrt{m/ \xi 1.7~10^{-4} eV}~\vec\beta_{earth} 
 \leq  (14/\sqrt{\xi}) \vec\beta_{earth}$,
  (considering that the mass
 of neutrinos with a very large lepton asymmetry is at most of the order of 0.1 eV) 
 and this term is dominant.

>From Eqs. (18) to (22) we have that in the rest frame of the electron, the frame 
in which experiments are performed, to first order in $\vec\beta_{earth}$ 
\begin{eqnarray}
H^D_{s_e} =&-&\frac{G_F}{\sqrt{2}} g_A 2\vec s_e\cdot\vec\beta_{earth}
\left[(n_{\nu}-n_{\nu^c})\right.\nonumber \\ 
&+&\left. \left(x\langle|\vec\beta_\nu|\rangle^{-1} + 
y\langle|\vec\beta_\nu|\rangle\right)
(n_{\nu_\ell} + n_{\nu^c_\ell}-n_{\nu_r}-n_{\nu^c_r}) \right]~
\label{eq:18} 
\end{eqnarray}
where $y$=1 and $x=0$ for relativistic neutrinos and $y$=2/3 and $x=16/3\pi=1.7$ for 
non-relativistic neutrinos (these numbers are the
prefactors in Eqs. (\ref{eq:197}), (\ref{eq:196}), and (\ref{eq:658}) above).
While in the rest frame of the neutrino bath, where $\langle\vec\beta_\nu\rangle = 0$ 
and $\vec\beta_e=\vec\beta_{earth}$ we find
\begin{equation}
H^D_{s_e} =  -\frac{G_F}{\sqrt{2}}g_A 2\vec 
s_e\cdot\vec\beta_{earth}\left[(n_{\nu}-n_{\nu^c}) +
\langle|\vec\beta_\nu|\rangle
(n_{\nu_\ell} + n_{\nu^c_\ell}-n_{\nu_r} - n_{\nu^c_r})\right].
\label{eq:200}
\end{equation}

For relativistic neutrinos, for which chirality and helicity coincide, we find 
in both 
reference frames that $H^D_{s_e} = H^D_R$ is
\begin{equation}
H^D_R =  -\sqrt{2}G_F g_A ~2\vec s_e\cdot\vec\beta_{earth}
(n_{\nu_L}-n_{\nu^c_R})=-\sqrt{2}G_F g_A ~2\vec 
s_e\cdot\vec\beta_{earth}(n_\nu-n_{\nu^c})~,
\label{eq:19} 
\end{equation}
and for non-relativistic gravitationally bound (C-NR) Dirac neutrinos (with well mixed 
helicities)   we find 
to order $\vec\beta_{earth}$, again in both 
reference frames, that  $H^D_{s_e} = H^D_{C-NR}$ is
\begin{equation}
H^D_{C-NR}= -\frac{G_F}{\sqrt{2}}g_A ~2\vec s_e\cdot\vec\beta_{earth}
(n_{\nu}-n_{\nu^c})=\frac{1}{2}H^D_R~,
\label{eq:195}
\end{equation}
which is smaller than the effect in Eq.~(\ref{eq:19}) for relativistic neutrinos 
by a factor of two.

For most of the non-relativistic relic neutrinos, those non-clustered, in the presence 
lepton asymmetry,  we find that in the rest frame of the electron the dominant
 contribution to $H^D_{s_e} = H^D_{NC-NR}$ is
\begin{equation}
H^D_{NC-NR}\simeq -\frac{G_F}{\sqrt{2}} g_A 2\vec s_e\cdot\vec\beta_{earth}
 1.7 \langle|\vec\beta_\nu|\rangle^{-1} 
(n_{\nu_\ell} + n_{\nu^c_\ell}-n_{\nu_r}-n_{\nu^c_r})~,
\label{eq:195new}
\end{equation}
with $\langle|\vec\beta_\nu|\rangle^{-1}$ = $\sqrt{m/ \xi 1.7~10^{-4} eV}~\vec\beta_{earth} 
 \leq  (8.2/\sqrt{\xi}) \vec\beta_{earth}$.

It is obvious that the effect is non-zero only in the presence  of a lepton 
asymmetry, where  $n_\nu \not= n_{\nu^c}$.  The effect is maximum if the
relic bath consists only of neutrinos (or only of antineutrinos) 
so that $n_\nu\not=0$ and 
$n_{\nu^c}=0$ (or viceversa), which is possible
with a large lepton asymmetry ($\mu/T > 2$). 
In this case, Eq.~(\ref{eq:19}) becomes
\begin{equation}
H^D_R = - \sqrt{2}G_F g_A 2\vec s_e\cdot\vec\beta_{earth}~n_\nu~,
\label{eq:20} 
\end{equation}
 and Eq.~(\ref{eq:195new}) becomes
\begin{equation}
H^D_{NC-NR}\simeq 0.85 \sqrt{\frac{m_\nu}{\xi 1.7 ~10^{-4}eV}}
 H^D_R \leq \frac{7}{\sqrt{\xi}} H^D_R~,
\label{eq:20new} 
\end{equation}
since $m_\nu \leq 0.1 eV$, in the presence of a large lepton asymmetry.

\subsection{Majorana Neutrinos}

In comparison to Dirac neutrinos, Majorana neutrinos satisfy an additional
constraint, $\nu = \nu^c$.  We write the 
Majorana field as 
$\Psi^M_\nu = (\Psi^D_\nu + (\Psi^D_\nu)^c)/\sqrt{2}$.  Therefore, using 
the ordinary decomposition for Dirac neutrinos $\Psi^D_\nu$ one arrives at
\begin{eqnarray}
\Psi^M_\nu(x,t) = \int \frac {d^3k}{(2\pi)^{3/2}} \left(\frac 
{m_\nu}{E_\nu}\right)^{1/2} 
\sum_{s} 
\left[\left(\frac{b(k,s) + d(k,s)}{\sqrt{2}}\right) u(k,s)e^{-ik\cdot 
x}\right.\nonumber\\
+ \left.\left(\frac{b(k,s) + d(k,s)}{\sqrt{2}}\right)^{\dagger}v(k,s)e^{ik\cdot 
x}\right]
\label{eq:21new}
\end{eqnarray}
Defining a new operator $\tilde{b}(k,s)=(b(k,s)+d(k,s))/\sqrt{2}$ 
 the general wave expansion for a Majorana field is given in terms of only 
one
creation and one annihilation operator, $\tilde{b}$ and $\tilde{b}^{\dagger}$.  
Note that the factor $1/\sqrt{2}$ is included in the definition
of $\tilde{b}$ and $\tilde{b}^{\dagger}$, instead of keeping it as an overall 
factor.  The 
overall normalization of a Majorana field is trickier than that of a Dirac 
field.
Our Dirac fields are normalized such that $\int 
d^3x{\Psi^D_\nu}^{\dagger}\Psi^D_\nu = 
N_\nu-N_{\bar\nu}$.
However, in the case of Majorana fields, $\int 
d^3x{\Psi^M_\nu}^{\dagger}\Psi^M_\nu=0$, and 
obviously such a
condition cannot be used to determine an overall normalization.  However, one 
can check the anti-commutation relations of the  creation and annihilation 
operators $\tilde{b}$ and $\tilde{b}^{\dagger}$.  One finds that
$\left\{\tilde{b}(k,s),\tilde{b}^{\dagger}(k,s^\prime)\right\}
=\delta_{s,s^\prime}\delta^3(\vec  k - \vec k^\prime)$,
which is the correct relation and proves  
that the definitions of $\tilde{b}$ and $\tilde{b}^{\dagger}$ 
are correct and that no further normalization
of $\Psi^M_\nu$ is needed.

Since the general wave expansion of the Majorana field is written in terms of 
only one creation and one annihilation operators, one can recover the  result for
Majorana neutrinos directly
from the calculation for  Dirac neutrinos by taking $b=d$ in Eq.~(\ref{eq:7}).
Effectively the vertex $\gamma_\mu(1 - \gamma_5)$ in Eq.~(\ref{eq:3}) becomes
a pure axial $\gamma_\mu\gamma_5$ vertex. This amounts
to the absence of the $k_\mu$ terms in Eq.~(\ref{eq:5}) and 
thus the terms proportional to $(n_\nu- n_{\nu^c})$ in  Eq.~(\ref{eq:16}) are 
absent. Therefore, the
effect depends on having  a net helicity in the Majorana-neutrino bath.  
The terms linear in
$\vec s_e$ in $H$ are now:
\begin{eqnarray}
H^M_{s_e} &=&  \sqrt{2} G_F g_A\left\{-2\vec s_e\cdot\vec\beta_e
\langle|\vec\beta_\nu|\rangle\right.\nonumber\\
&+& 2\vec s_e\cdot \left. \left(
\left<\frac{E_\nu}{E_\nu+m_\nu}\vec\beta_\nu|\vec\beta_\nu|\right>
+\left< \frac{m_\nu}{E_\nu}~\hat\beta_\nu \right> \right) \right\}
(n_{\nu_\ell}-n_{\nu_r})~.
\label{eq:21} 
\end{eqnarray}
In the rest frame of the electron, the frame relevant for experiments,
this is
\begin{equation}
H^M_{s_e} = -\sqrt{2} G_F g_A ~2\vec s_e\cdot\vec\beta_{earth}
\left[x\langle|\vec\beta_\nu|\rangle^{-1} + y \langle|\vec\beta_\nu|\rangle 
\right]
~(n_{\nu_\ell}-n_{\nu_r})
\label{eq:22} 
\end{equation}
where, as before, $x$ = 0 and $y$=1 for relativistic and $x$=$16/3\pi$ and 
$y$=2/3 for 
non-relativistic neutrinos.
For relativistic Majorana neutrinos, for which helicity coincides with
chirality $\nu_\ell=\nu_L$ and $\nu_r=\nu^c_R$, this term coincides with the 
result for relativistic
Dirac neutrinos in Eq.~(\ref{eq:19}), as it should be since Dirac and 
Majorana neutrinos cannot be distinguished when relativistic.  
In the absence of interactions the helicity is conserved, even when as the 
temperature
of the universe decreases neutrinos become non-relativistic.
In the case of a large lepton asymmetry favoring, say, neutrinos, so that
only $\nu_L=\nu_\ell$ are present at the moment of decoupling (when neutrinos 
are relativistic) a net left helicity  remains in the bath of Majorana
neutrinos, while not gravitationally bound.
 In this case,  without
gravitational binding, Majorana neutrinos would have a net left helicity
$n_{\nu_\ell}=n_{\nu_L}\not= 0$, $n_{\nu_r} = n_{\nu_R^c}=0$. 
 As in the case of Dirac neutrinos the term proportional to 
$\langle|\vec\beta_\nu|\rangle^{-1}$ dominates for non relativistic neutrinos.
This is important only for unclustered  (NC-NR) neutrinos.
As already mentioned, gravitationally bound non-relativistic (C-NR)
neutrinos loose their net helicity.  After a characteristic orbital time,
 helicities become maximally 
mixed  thus  $\ell$ and $r$ helicities become
equally abundant, so $(n_{\nu_\ell}-n_{\nu_r})=0$, and
the Stodolsky effect vanishes.

Thus, to summarize, in general
\begin{equation}
H^M_R = H^D_R
\label{eq:193}
\end{equation}
\begin{equation}
H^M_{C-NR} = 0 
\label{eq:192}
\end{equation}
and for neutrinos with a large lepton asymmetry, so that say $n_{\nu_\ell} \simeq n_\nu$, 
\begin{equation}
H^M_{NC-NR}\simeq 1.7 \sqrt{\frac{m_\nu}{\xi 1.7 ~10^{-4}eV}}
 H^D_R \leq \frac{14}{\sqrt{\xi}} H^D_R~,
\label{eq:192new} 
\end{equation}
In this last equation $H^D_R$ is that given in Eq.~(\ref{eq:20}).  

\subsection{Accelerations}

The Stodolsky effect consist of a  difference in energy between 
the two helicity states of the electron interacting with the neutrino bath in the rest 
frame of the earth  in which experiments are performed (which we take to coincide with 
the rest frame of the electrons). This energy difference $\Delta E$ is obtained in each 
case by replacing $\vec s_e\cdot\vec\beta_{earth}$
 by $|\vec\beta_{earth}|$ in the respective expressions for the $H^D$ and $H^M$
given in the previous sections.  Therefore, the Stodolsky effect vanishes except in the 
presence of a lepton asymmetry which in the case of Majorana neutrinos persists in the 
neutrino bath as an asymmetry in helicity as long as neutrinos are not gravitationally 
bound.
In the case of relativistic, light neutrinos of density $n_\nu$, with a very 
large lepton asymmetry favoring, say, neutrinos $\nu_L$, so that 
$n_{\nu_L}=n_{\nu_\ell}=  n_\nu$, from Eqs.~(\ref{eq:19})
 and (\ref{eq:193}) one has 
\begin{equation}
(\Delta E)^D_R = (\Delta E)^M_R
=2\sqrt{2} G_Fg_A~
|\vec\beta_{earth}|n_\nu~.
\label{eq:25} 
\end{equation}
In the case of non-relativistic, unclustered neutrinos, we see from Eqs.~(\ref{eq:20new})
 and (\ref{eq:192new}) that 
\begin{equation}
(\Delta E)^M_{NC-NR}= 2 (\Delta E)^D_{NC-NR}
\simeq 1.7 \sqrt{\frac{m_\nu}{\xi 1.7 ~10^{-4}eV}}
 (\Delta E)^D_R \leq \frac{14}{\sqrt{\xi}} (\Delta E)^D_R~.
\label{eq:25new} 
\end{equation}
In the case of gravitationally bound (C-NR) neutrinos (so that 
there is no net helicity in the bath) of density $n_\nu$, if there is a very large 
lepton asymmetry favoring neutrinos  $\nu_L$, 
for Dirac neutrinos $n_{\nu^c_R}=0$, and for Majorana neutrinos 
$n_{\nu_{\ell}}=n_{\nu_r}$
and the effect vanishes.  Thus,
\begin{equation}
(\Delta E)^ D_{C-NR} \simeq \sqrt{2} G_F 
g_A|\vec\beta_{earth}|n_{\nu}~,
\label{eq:26} 
\end{equation}
\begin{equation}
(\Delta E)^M_{C-NR} \simeq 0~.
\label{eq:27} 
\end{equation}
Equivalent results would be obtained with an asymmetry favoring antineutrinos.

\vspace{5 mm}

>From now on, we will use the energy difference $\Delta E$ in Eq. (\ref{eq:25}) 
to compute the maximal possible strength of the Stodolski effect, recalling 
that the effect could be at most one order of magnitude larger for 
non-clustered non-relativistic (NC-NR) neutrinos (most of the relic neutrinos
if they are non-relativistic).

\vspace{5 mm}

The difference in energy $\Delta E$ between the two helicity states 
of an electron in the direction of the bulk velocity
of the neutrino background $<\vec\beta_\nu>=-\vec\beta_{earth}$ implies a torque 
of magnitude
$\Delta E/\pi$  applied on the spin of the electron.  Since the spin is
``frozen" in a magnetized macroscopic piece of material with $N$ polarized
electrons,
the total torque applied to the piece has a magnitude  $\tau = N\Delta E/\pi$.
Given a linear dimension $R$ and mass $M$ of the macroscopic object, its
moment of inertia is parametrized as $I = MR^2/\gamma$, where $\gamma$ is a 
geometrical factor.  In the typical case of one polarized electron per atom in a material 
of atomic number $A$, the number $N$ above is $N = (M/gr) N_{\rm AV}/A$ (using cgs units), 
where $N_{\rm AV}$ is Avogadro's number. 
Thus, the effect we are considering would produce an
angular acceleration of order $\alpha = \tau/I$ and 
a linear accelerations of order $a_{G_F} = R\alpha$ in the magnet, given by
\begin{equation}
a_{G_F} = \frac{N_{\rm AV}}{A}\frac{\Delta E}{\pi}\frac{\gamma}{R~(gr)}
\label{eq:28} 
\end{equation}
where the $G_F$ subindex indicates the mechanism we have described.

Using the  expression in  Eq. (\ref{eq:25}) for $\Delta E$  with 
$n_\nu = f~100 cm^{-3}$ we then find
\begin{equation}
a_{G_F}  =10^{-27} f\cdot\left(\frac{\gamma}{10}\right)
\left(\frac{100}{A}\right)\left(\frac 
{cm}{R}\right)\left(\frac{\beta_{earth}}{10^{-3}}\right)~
\frac{\rm cm}{{\rm sec}^2}~
\label{eq:29} 
\end{equation}
where $f$ accounts for a possible local enhancement of the standard background
neutrino number density. This acceleration could be at most one order of magnitude
larger for non-relativistic (NC-NR). These accelerations are rather weak.  

As mentioned above the present upper bound from structure formation in the 
Universe for neutrinos 
without a lepton  asymmetry, for which $n_\nu \simeq 100$ cm$^{-3}$, is of a few eV. 
The upper bound on the mass of degenerate neutrinos with a large asymmetry should be 
lower, since their number density is larger. Light 
neutrinos with masses $m_\nu < eV$ would be either unbound or
gravitationally bound to very large structures 
and thus local enhancement to the neutrino density 
due to gravitational clustering  would be very small.

Throughout we have used $\beta_{earth}\approx 10^{-3}$, which is 
the velocity of the earth with respect to the galaxy; however, if neutrinos are 
gravitationally clustered on much larger scales, 
one must consider the relative motion of the earth with respect to the rest frame of 
these larger objects.    At such scales the neutrino bath rest frame can be taken to
coincide with the cosmic microwave background (CMB) rest frame.  
The sun's motion with respect to the CMB
is believed to be responsible for the largest anisotropy in the COBE 
DMR maps, the 3 mK dipole, and thus has been determined
with great accuracy.  From the COBE data, $v_{sun}=369.0\pm2.5$ km/sec which 
corresponds to $\beta_{earth}=1.231\times 10^{-3}$ 
\cite{lineweaver}.  Therefore, even if background neutrinos are bound to such large 
objects as super-clusters, we are justified in using still
$\beta_{earth}\approx 10^{-3}$.  

Thus with very little gravitationally clustering locally, a neutrino density 
enhancement could only be due to the asymmetry itself, in which case (as explained in 
the introduction) $f < 20$ ($n_{\nu_i} \le 1700$ cm$^{-3}$).  

However, it is amusing to note that if there is a local
cloud of $\nu_e$ of density $\simeq 10^{17}$ cm$^{-3}$ as required
for a rather audacious explanation of the tritium end point 
anomaly\cite{Robertson} then 
$a_{G_F} \simeq 10^{-12}$ cm/sec$^2$, 
and the effect would be observable!

\subsection{Comparisons with Past Results}

The first calculation of this order $G_F$ effect was done by Stodolsky in 1974 
\cite{Stodolsky}.  
In our notation, he found the energy difference between the two
spin states of an electron moving with velocity $\vec\beta_e$ in the rest frame 
of the 
neutrino bath
to be 
\begin{equation}
\Delta E=2\sqrt{2}G_F \frac{\beta_e}{\sqrt{1-\beta^2_e}}\left(n_\nu - 
n_{\nu^c}\right)~.
\label{eq:101}
\end{equation}
Stodolsky performed his calculation before the discovery of neutral currents and 
thus
did not include them (hence he took $g_A$ to be 1).  He considered only 
relativistic
neutrinos.  Stodolsky's result conincides with ours in Eq.~(\ref{eq:25}) with
$\beta_e=|\vec\beta_{earth}|$ (and $\sqrt{1-\beta^2_e}\approx 1$).

Langacker, Leveille, and Sheiman also explored this effect \cite{Langacker}.  
Although
the majority of their paper deals with exposing the flaws in several relic 
neutrino   
detection schemes, they conclude that the only legitimate effect of order $G_F$ 
is that
proposed by Stodolsky.  They studied only Dirac neutrinos and found
\begin{equation}
\Delta E=2\sqrt{2} G_F \beta_e N^{tot}_a \sum_i \left(n_{\nu_{i}} - 
n_{\nu^c_i}\right)g^{ai}_A K(p,m_i)~,
\label{eq:102}
\end{equation}
where $N^{tot}_a$ is the total number of electrons in the sample, and K is a 
function of
$E_\nu$ and $m_\nu$ which reduces to 1 if $m_\nu=0$ or 1/2 if $k_\nu \ll m_\nu$. 
 Their 
calculation
is performed in the rest-frame of the ``neutrino sea".  
Their result agrees exactly with Eqs.~(\ref{eq:25}) and (\ref{eq:26}) in which we
give the $\Delta E$ for Dirac neutrinos in the rest frame of the electron.

Ferreras and Wasserman \cite{Ferreras} recently tackled the subject of the
detectability of relic neutrinos, and also concluded that there are no order 
$G_F$ effects 
without
lepton asymmetry.  However, they point out that density fluctuations in the the 
$\nu(\nu^c)$ 
background
could give rise to non-steady forces of order $G_F$.  Such forces would give 
rise to 
displacements
in massive objects proportional to $t^{3/2}$ rather than $t^2$ as in constant 
acceleration.  
They caution
that although such accelerations may be more readily detectable, they could also 
act as an 
additional source
of noise.  Here we have not considered neutrino density fluctuations.

\section{The $G_F^2$ Effect}

Nucleons are continuously bombarded by the relic neutrinos and a momentum 
$\Delta\vec p \approx \vec p_\nu$ is
imparted in each collision.  The momentum of the neutrinos $p_\nu$ is 
$p_\nu=E_\nu/c \approx 4T_\nu/c$ for 
relativistic neutrinos (R) and also for non-clustered, non-relativistic 
neutrinos (NC-NR) due to momentum red-shift,
and is $m_\nu v_\nu \approx m_\nu v_{virial}$ for non-relativistic clustered 
neutrinos (C-NR). 

If the earth was at rest with respect to the relic neutrino ``rest frame", i.e. 
the frame in which the neutrinos are isotropically distributed, then the average 
momentum transfer $\left<\Delta p\right>$
would vanish.  However the motion of the earth with velocity 
$c\beta_{earth}=v_{earth}$ induces a 
``dipole" distortion of $O(\beta_{earth})$ in the velocity distribution of the relic 
neutrinos.   In the laboratory frame (the earth's rest 
frame) this makes 
\begin{equation}
\begin{array}{ll}
\left<\Delta p\right>_R \approx \beta_{earth} \left(E_\nu/c\right)~, \\
\left<\Delta p\right>_{NC-NR} \approx \beta_{earth} \left(4T_\nu/c\right) = 
\left<\Delta 
p\right>_R~, \\
\left<\Delta p\right>_{C-NR} \approx \beta_{earth}~c~m_\nu~.
\end{array}
\label{eq:1000}
\end{equation}

The fluxes of infalling neutrinos are $\Phi_R = n_\nu c$ and $\Phi_{NR}=n_\nu 
v_\nu$ for 
relativistic and 
non-relativistic neutrinos respectively.  To find the resulting accelerations we 
compute the 
force exerted
on one gram of detector material containing $\mathcal{N}=$$N_{AV}/A$ nuclei 
(A,Z).  The force 
is given by the 
momentum imparted per second: the latter is the microscopic $\Delta p$ of Eq. 
(\ref{eq:1000}) 
times 
$\mathcal{N}$$\Phi_\nu \sigma_{\nu-A}$, the number of collisions per second 
inside one gram:

\begin{equation}
\frac{F}{m} = a = \Phi_\nu \frac{N_{AV}}{A} \sigma_{\nu-A} \langle \Delta 
p\rangle~.
\label{eq:1001}
\end{equation}

The neutrino-nucleus cross sections $\sigma_{\nu-A}$ are extremely small even if 
we include a 
nuclear coherence
enhancement factor $\left(A-Z\right)^2\approx A^2$
relative to the neutrino-nucleon $\sigma_{\nu-N}$ cross sections, which are of 
the order

\begin{equation}
\sigma_{\nu-N} \approx \left\{\begin{array}{ll}
G_F^2 m_\nu^2/\pi \simeq 10^{-56}\left({m_\nu}_{\{\rm eV\}}\right)^2~
{\rm cm^2} & {\mbox{for (NR)}} \\
G_F^2 E_\nu^2/\pi \simeq 5 \times 10^{-63}\left({T_\nu}_{\{1.9^{\circ} K\}} 
\right)^2~
{\rm cm}^2 & {\mbox{for (R)}}
\end{array} \right.
\label{eq:1002}
\end{equation}

where $\{~\}$ indicates the units of the mentioned quantities.

Collecting all terms in Eq. (\ref{eq:1001}) above we find

\begin{equation}
a=\frac{N_{AV}}{A}n_\nu \frac{{G_F}^2}{\pi} A^2 \left\{
\begin{array}{lll}
\left(4 T_\nu\right)^3 \beta_{earth} & {\mbox{for (R)}} \\
{m_\nu}^2 4 T_\nu \beta_{earth} & {\mbox{for (NC-NR)}}~~. \\
{m_\nu}^3 {v_\nu}^2 & {\mbox{for (C-NR)}} 
\end{array} \right.
\label{eq:1003}
\end{equation}

We will take the earth's velocity $v_{earth}$ and the clustered neutrino's 
velocity
to be all $\approx c\beta_{virial}$ with $\beta_{virial}\approx 10^{-3}$ 
corresponding
to typical galactic virial velocities.  We use $n_\nu= f$ 100 cm$^{-3}$.  Hence 
we find

\begin{equation}
a_R = 3\times10^{-54} f\frac{A}{100}\left({T_\nu}_{\{1.9^{\circ} K\}}\right)^3 
\frac{cm}{sec^2}~,
\label{eq:1004}
\end{equation}

\begin{equation}
a_{NC-NR}=0.6\times10^{-47} f\frac{A}{100}\left({m_\nu}_{\{ev\}}\right)^2
{T_\nu}_{\{1.9^{\circ} K\}} \frac{cm}{sec^2}~.
\label{eq:1005}
\end{equation}

Here the density enhancement factor $f$ may only be due to a large lepton 
asymmetry and 
$f<20$ 
(see Sect. 1).  Finally for clustered non-relativistic neutrinos we have

\begin{equation}
a_{C-NR}=10^{-46}f\frac{A}{100} \left({m_\nu}_{\{ev\}}\right)^3 
\frac{cm}{sec^2}~.
\label{eq:1006}
\end{equation}

Here f contains also a clustering enhancement factor.  Values of $f$ up to 
$10^7$ have been
mentioned in the literature (see for example \cite{Hagmann}).  If neutrinos were 
sufficiently massive to cluster in our galaxy and make up the local dark matter 
halo ($m_\nu \ge 20 eV$),
only then would we have $f=\rho_{local}/m_\nu=0.4\times 10^7/{m_\nu}_{\{eV\}}$.  
This possibility
is already rejected by structure formation arguments.  With sub-eV neutrino 
masses the enhacement f due to clustering could 
only be of close to 1.

All of the above accelerations are extremely small and beyond the reach of any 
known  experimental measurement technology.  
However, it has been noted (by Zeldovich and Khlopov 
\cite{Zeldovich} as
well as Smith and Lewin \cite{SmithLewin}) that coherent scattering of neutrinos 
from domains
of the size of the deBroglie neutrino wavelength $\lambda_\nu=2\pi \hbar/p_\nu$ 
dramatically 
increases the scattering cross section.
The extra factor due to coherence is the number of nuclei in this domain (since 
the nuclear 
coherence
factor is already included),

\begin{equation}
N_c=\frac{N_{AV}}{A}\rho_{\{gr~cm^{-3}\}} \left(\lambda_{\nu 
{\{cm\}}}\right)^3~.
\label{eq:1008}
\end{equation}

For relativistic and for unclustered non-relativistic relic neutrinos with 
$\lambda_\nu=2\pi\hbar/4T_\nu$ $\approx$ 0.2 cm this enhancement factor is

\begin{equation}
N_c = \frac{5\times 10^{21}}{A~{{T_\nu}_{\{1.9^{\circ} K\}}}^3} 
\rho_{\{gr~cm^{-3}\}}~.
\label{eq:1009}
\end{equation}

For clustered non-relativistic neutrinos with $p_\nu=m v_\nu \approx 10^{-3} 
{m_\nu}_{\{eV\}}$ $eV$ 
the wavelength is $\lambda_\nu \approx \left(0.12/{m_\nu}_{\{eV\}}\right)$ cm so 
that the 
coherence enhancement
factor is 

\begin{equation}
N_c= \frac{10^{21}}{A~{{m_\nu}_{\{ev\}}}^3}  \rho_{\{gr~cm^{-3}\}}~.
\label{1007}
\end{equation}

Thus the largest acceleration values with the $N_C$ enhancement factors are

\begin{equation}
a_R = 2\times10^{-34} f \rho_{\{gr~cm^{-3}\}} \frac{cm}{sec^2}~,
\label{eq:2000}
\end{equation}

\begin{equation}
a_{NC-NR}=3\times10^{-28} f \left({m_\nu}_{\{ev\}}\right)^2
\left({T_\nu}_{\{1.9^{\circ} K\}}\right)^{-2} \rho_{\{gr~cm^{-3}\}} 
\frac{cm}{sec^2}~,
\label{eq:2001}
\end{equation}

and finally, for clustered, non-relativistic neutrinos
 
\begin{equation}
a_{C-NR} \approx 10^{-27} f \rho_{\{gr~cm^{-3}\}} \frac{cm}{sec^2}~.
\label{eq:2002}
\end{equation}

The above discussion applies only to Dirac non-relativistic neutrinos as only 
these have
coherent vectorial $Z^o$ couplings in the static limit
\begin{equation}
A_{Z^o}^{(o)}\cdot \bar\psi\gamma_o\psi\approx
A_{Z^o}^{(o)}\psi^+\psi~.
\label{eq:1011}
\end{equation}
In the Majorana case we still have coherent $\psi^+\gamma_5\psi A_{Z^o}^{(o)}$
couplings.  The latter are, however, suppressed by $\beta_\nu$, the ratio
of ``small" and ``large" components of the spinor.  Recall that for 
non-relativistic spinors the lower components are ``smaller'' than the upper components 
by a factor of $\beta$.  Without
the $\gamma_5$ these terms are dwarfed by the leading order terms;  however with 
the $\gamma_\mu \gamma_5$
vertex in the static limit these terms are the leading order terms).   Thus the 
analog of Eqs. (\ref{eq:2001}) and (\ref{eq:2002}) for
non-relativistic Majorana neutrinos is suppressed by an extra factor of 
$\beta_\nu^2\approx 10^{-6}$.

\subsection{Solar Neutrinos and WIMPS}

Solar neutrinos provide a directional  fairly well known source of 
relativistic neutrinos, and it is interesting to estimate their contribution to the 
accelerations we have calculated.  The acceleration due to solar neutrinos is
\begin{equation}
a = \Phi_{{\rm solar}~\nu} ~ \frac{N_{AV}}{A} \rho \sigma_{(\nu-A)}
\frac{E_\nu}{c}~.
\label{eq:1012}
\end{equation}
For a galium detector, for example,  we take $\Phi_\nu \simeq 10^{11} {\rm 
cm}^{-2}{\rm 
sec}^{-1}$, $E_\nu \simeq 0.3$ MeV,  and $\sigma_{\nu-{\rm galium}} \approx 
10^{-44}$
cm$^2$.  Thus for solar pp neutrinos in galium we find
\begin{equation}
a  \approx 10^{-27}~{\rm cm/sec^2}.
\label{eq:1013}
\end{equation}
This exceeds most previously calculated accelerations, but it is still too small 
to be 
detectable
at present.

Our main focus here is on neutrino induced forces.  However, it is believed that 
much of
the cosmological and most of the halo dark matter is made of massive weakly
interacting particles (WIMPs).  

Let us first estimate the effect of  WIMPs if they had the cross section of Dirac 
neutrinos with masses $m_X\approx$ O(100 GeV). Then the nuclear cross section with 
$A\simeq 100$ targets would be large
\begin{equation}
\sigma_{x-A} \approx \frac{G_F^2}{\pi} \left(m_X\right)^2
A^2 \approx 10^{-30}{m_X}_{\{GeV\}}A^2~{\rm cm}^2
\label{eq:1014}
\end{equation}
and the recoil energies, $m_X \beta_X^2/2 ~\approx O(30~{\rm 
keV})$ (using $m_X\approx 100 {\rm GeV}$), detectable.  Indeed such WIMPs 
have been excluded by direct searches in underground detectors. Realistic WIMP 
candidates at the  ``threshold of
detectability" have smaller  cross-sections by a factor $10^{-\Delta}$.  
 Using the analog of Eq. (\ref{eq:1001})
\begin{eqnarray}
a_{\rm WIMP} &=& \Phi_{\rm WIMP} \frac{N_{AV}}{A} \sigma_{(X-A)} m_X 
v_X\nonumber \\
&=& n_X m_X \frac{N_{AV}}{A} \sigma_{(X-A)} {v_X}^2,
\label{eq:1015}
\end{eqnarray}
and a WIMP density 
$n_X m_X\simeq\rho_{\rm dark(local)} \simeq 10^{-24}~{\rm gr/cm}^3$, we
find for $A=100$ and $\sigma_{X-A} \approx 10^{-30-\Delta}$ cm$^2$
\begin{equation}
a_{\rm WIMP} \approx 10^{-(17 + \Delta)}
\left(\frac{\beta_X}{10^{-3}}\right)^2\frac{A}{100}~ \frac{{\rm cm}}{{\rm 
sec}^2}~.
\label{eq:1016}
\end{equation}

Clearly $a_{\rm WIMP}$ dominates over a very large range of cross sections 
(with $\Delta < 10$, using Eq. (39) or Eq. (54)) the corresponding accelerations 
due to the scattering of light, locally unclustered
neutrinos.

\section{Conclusions}

We have calculated the magnitude of the signals expected for realistic cosmic 
neutrino backgrounds in detectors attempting to measure the mechanical forces exerted 
on macroscopic targets by the elastic scattering of relic neutrinos.  We examined 
effects proportional to $G_F$and $G_F^2$ for both Dirac and Majorana neutrinos either 
relativistic or non-relativistic.
We also estimated the contributions to macroscopic accelerations due to solar 
neutrinos and WIMPs in the galactic halo.

The effect linearly proportional to the weak coupling constant $G_F$
vanishes in the case of no lepton asymmetry.  With a lepton 
asymmetry, macroscopic accelerations for relativistic Dirac 
and Majorana neutrinos and
clustered non-relativistic Dirac neutrinos were found in Eq. (\ref{eq:29}) 
to be  of the order of $f \cdot 10^{-27}$ cm/sec$^2$, 
where f is a density enhancement
factor  which  can be at most about 17 (with a large lepton asymmetry), 
 and at most one order
 of magnitude larger (i.e. $f \cdot 10^{-26}$ cm/sec$^2$)
 for non-clustered Dirac or Majorana neutrinos (which are most of the relic neutrinos
if they are non-relativistic at present). These 
accelerations are at most thirteen orders of magnitude smaller that
 the smallest measureable 
acceleration of  $10^{-12}$ cm/sec$^2$. 
The acceleration of non-relativistic, 
gravitationally bound Majorana neutrinos vanishes.  

For the effect proportional to $G_F^2$ accelerations of relativistic Dirac and 
Majorana neutrinos, taking advantage of coherent scattering effects, were found in Eq. 
(\ref{eq:2000}) to be of the order
of $f \cdot 10^{-34}$ cm/sec$^2$.  The accelerations of non-relativistic Dirac 
neutrinos were calculated  to
be of the order of $f \cdot 10^{-28} (m_\nu/$eV)$^2$ cm/sec$^2$ 
in Eq. (\ref{eq:2001}) for 
non-clustered neutrinos  and $f \cdot 10^{-27}$ 
cm/sec$^2$ in Eq. (\ref{eq:2002}) for clustered neutrinos, while the accelerations of 
non-relativistic Majorana  neutrinos
are down by a factor $\beta_\nu^2 \simeq 10^{-6}$.  All 
accelerations are well beyond the smallest measureable acceleration, $10^{-12}$ 
cm/sec$^2$, mentioned above.  

Additional calculations for the accelerations due to solar neutrinos and WIMPs 
in the galactic halo raise concerns that signals in a detector due to relic neutrinos 
may well be washed out by solar neutrino
or WIMP events unless directionality can be used to reject them.

\section*{Acknowledgements}

We thank Peter Smith, Leo Stodolsky and Joel Primack for clarifying discussions.
S.N. thanks the  Israeli National Science Foundation Grant No. 561/99 for 
support and UCLA 
for hospitality. G.D. and G.G. were supported in part by the U.S. Department of 
Energy Grant No. DE-FG03-91ER40662, Task C.


\begin{thebibliography}{9}

\bibitem{Stodolsky} L. Stodolsky, Phys. Rev. Lett. {\bf 34}, 110 (1974).

\bibitem{SmithLewin} P. Smith and J. Lewin, Phys. Lett. {\bf 127B}, 185 (1983).

\bibitem{Zeldovich2} V. Shvartsman, V. Braginski, S. Gershtein, Y. Zel'dovich, 
and M. 
Khlopov, JETP Lett. 
{\bf 36}, 277 (1982).

\bibitem{Langacker} P. Langacker, J. Leveille, and J. Sheiman, Phys. Rev. D {\bf 
27}, 1228 
(1983).

\bibitem{Ferreras} I. Ferreras and I. Wasserman, Phys. Rev. D {\bf 52}, 5459 
(1995).

\bibitem{Hagmann} C. Hagmann, astro-ph/9902102,
published in *Asilomar 1998, Particle physics and the early universe* 460-463. 

\bibitem{minacc} E. Adelberger et al., Annu. Rev. Nucl. Part. Sci.
{\bf 41}, 269 (1991).

\bibitem{KS} H. Kang and S. Steigman, Nucl. Phys. B{\bf 372}, 494 (1992).

\bibitem{AS} J. Adams and S. Sarkar, talk presented at the ``ICTP Workshop
on the Physics of Relic Neutrinos", Trieste, Italy, November 1998.

\bibitem{LP} J.~Lesgourgues and S.~Pastor,
Phys.\ Rev.\ D {\bf 60}, 103521 (1999).


\bibitem{KR} W.~H.~Kinney and A.~Riotto,
Phys.\ Rev.\ Lett.\  {\bf 83}, 3366 (1999).


\bibitem{S2} J.~P.~Kneller, R.~J.~Scherrer, G.~Steigman and T.~P.~Walker,
astro-ph/0101386.

\bibitem{primack} See, for example, J.~R.~Primack and M.~A.~Gross,
astro-ph/0007165.

\bibitem{tremaine-gunn} S. Tremaine and J. Gunn, Phys. Rev. Lett {\bf 42}, 407 
 (1979).
 
\bibitem{GurnseyDeWitt} G\"{u}rsey, F. (ed.), {\it Group Theoretical Concepts 
and Methods
in Elementary Particle Physics}, Gordon and Breach, (1964).  Refenced in {\it 
Quantum Field
Theory} by L.H. Ryder, p. 63. 

\bibitem{lineweaver} C.H. Lineweaver et. al., Ap. J. 470, 38 (1996).

\bibitem{Robertson} R.G.H. Robertson, T.J. Bowles, G.J. Stephenson Jr., D.L. 
Wark, J.F. 
Wilkerson, and D.A.
Knapp, Phys. Rev. Lett. {\bf 67}, 957 (1991).

\bibitem{Zeldovich} Y. Zeldovich and M. Khlopov, Sov. Phys. Usp. {\bf 24}, 755 
(1981).

\end{thebibliography}
\end{document}